\documentclass[fleqn,twoside]{article}
\usepackage{espcrc2}

\usepackage{graphicx}

\newcommand\cbar{{\overline{c}}}
\newcommand\Cbar{{\overline{C}}}
\newcommand\tr{{\rm tr}}
\newcommand\half{\frac{1}{2}}

\newcommand{\AmS}{{\protect\the\textfont2
  A\kern-.1667em\lower.5ex\hbox{M}\kern-.125emS}}

\title{$SU(N)$ chiral gauge theories on the lattice: a quick
overview}

\author{Maarten Golterman
        \address{Department of Physics and Astronomy,
        San Francisco State University,
        San Francisco, CA 94132, USA}%
        \thanks{Presenter at conference}
	and
        Yigal Shamir
        \address{School of Physics and Astronomy, Tel-Aviv University,
        Ramat Aviv, 69978 Israel}}

\begin{document}

\begin{abstract}
We describe how an $SU(N)$ chiral gauge theory can be put on
the lattice using non-perturbative gauge fixing.  In particular,
we explain how the Gribov problem is dealt with.  Our
construction is local, avoids doublers, and weak-coupling
perturbation theory applies at the critical point which defines
the continuum limit of our lattice chiral gauge theory.
\vspace{1pc}
\end{abstract}

\maketitle

The construction of lattice chiral gauge theories (ChGTs) is
an old problem.  Because of the Nielsen--Ninomiya theorem
\cite{nn} and the chiral anomaly \cite{ks}, one either has to give
up on chiral symmetry on the lattice, but enforce it to re-emerge
in the continuum limit, or modify the lattice definition of
chiral symmetry in order to maintain exact chiral symmetry on the
lattice. For reviews, see refs.~\cite{reviews,mgreview}.
Here we follow the first approach,
reporting on our recent completion of a construction based
on non-perturbative gauge fixing \cite{gsnonab}.
We will have some comments on the second approach towards the end.

If gauge invariance is broken, the longitudinal modes of the
gauge field couple to the fermions.  If the dynamics of the longitudinal modes
is uncontrolled, their ``back-reaction"
changes the fermion spectrum from chiral to vector-like \cite{reviews}.
A key point here is that having a reasonable definition
of the fermion determinant for smooth gauge fields
does not solve the problem.

A solution is to construct a 
lattice theory with a
critical point whose universality class is described by the
perturbative expansion of the target continuum theory \cite{ysgs}.
A renormalizable gauge is mandatory \cite{rome}, so that, in spite of
the lack of gauge invariance of the regulated theory, we can
use the usual power-counting arguments together with Slavnov--Taylor identities
to construct the counter terms, which are finite in number.

\includegraphics*[width=16.5pc]{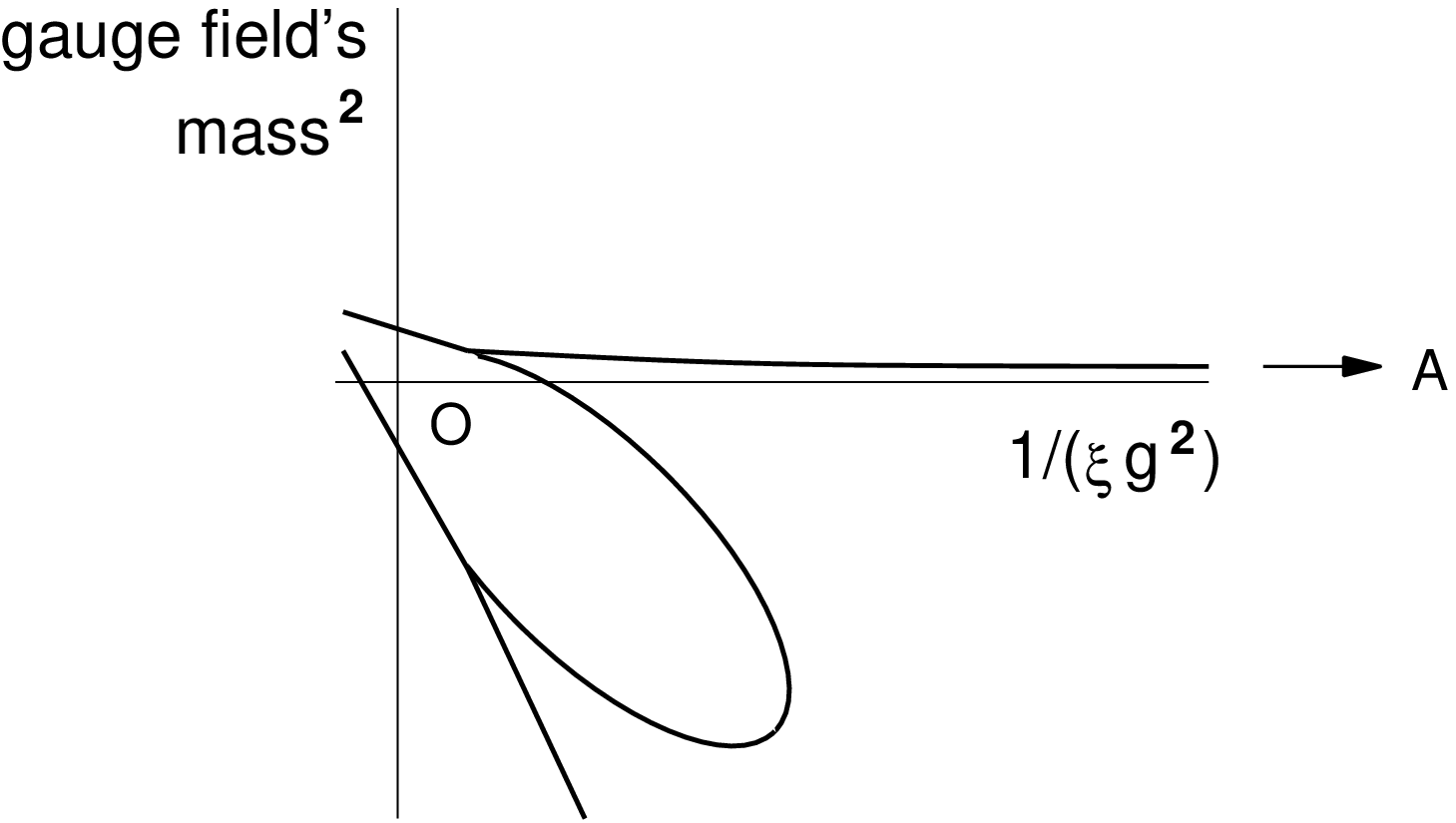}

Schematically, the critical point we are after is shown in the figure.
$g$ is the (bare) gauge coupling, and $1/\xi g^2$ is the
parameter multiplying the gauge-fixing action. The vertical axis
is the (relevant!) gauge field's mass-squared counter term.
(The axis corresponding to the bare coupling itself is not shown.)
The usual lattice Yang--Mills critical point is at the origin.
But here the longitudinal modes are uncontrolled,
and destroy the chiral nature of the fermion spectrum.
The critical point we aim for is point $A$,
which involves both the limit $g \to 0$ (for example, at fixed $\xi$) and
tuning the mass counter term to the critical line shown in the figure.
(Obviously, all other counter terms need to be tuned as well, see
below.)

In order to secure the desired critical point
and recover the target chiral gauge theory, we need
lattice perturbation theory (PT) to be valid near $A$.
Whether this can be accomplished or not is a non-trivial
non-perturbative question.  As such, it cannot be addressed
in the context of continuum perturbation theory, where it is
simply {\it assumed} that some non-perturbative theory exists for which
the usual perturbative expansion is valid.  In a non-perturbative context
one cannot
assume the existence of this critical point, but instead one has to
demonstrate that it exists.

Our actual construction of the appropriate critical point is
achieved by using a lattice gauge-fixing action that
1) contains a lattice transcription of the longitudinal kinetic term
$(\partial\cdot V)^2/2\xi g^2$ \cite{rome}, and
2) has the unique minimum $U_\mu=\exp(iV_\mu)=1$ \cite{ysgs}.
This guarantees sufficient control over all short-distance effects
(for details, see ref.~\cite{gsnonab}).
In particular, no fermion doublers are generated if they are not
present in the classical continuum limit \cite{bgsprl,bd}.

If a regularization breaks (chiral) gauge symmetry, counter terms
are needed to restore gauge invariance.
One adjusts these counter terms such that the renormalized theory
is invariant under BRST symmetry
(this is possible to all orders for an anomaly-free fermion spectrum).
In non-abelian theories,
this raises the question as to how BRST symmetry works non-perturbatively.
Our new work \cite{gsnonab} answers this question.

We begin with reviewing the situation.  One starts with
$Z=\int dU\,\exp{[-S_{inv}(U)]}$, an un-fixed
theory on the lattice with compact gauge fields. Following Faddeev--Popov,
one inserts
\begin{eqnarray}
Z_{gf}&\!\!=\!&\int d\phi dc d\cbar db\,\exp{[-S_{gf}(U^\phi,c,\cbar,b)]}\,,\\
S_{gf}&\!\!=\!&2t[-i\tr(bF(U))+tr(\cbar{\cal M}c)]+\xi g^2\tr(b^2)\,,
\nonumber
\end{eqnarray}
with $F(U)$ the gauge-fixing condition
and ${\cal M}c = \delta_{BRST} F(U)$.
The requirement is that $Z_{gf}$ is independent of $U$,
so that only a constant was inserted into $Z$.  If this holds, the
gauge-invariant correlation functions of the gauge-fixed theory are
identical to those of the un-fixed theory.

The good news is that, thanks to BRST invariance, $dZ_{gf}/dt=0$ \cite{nnogo}.
Hence $Z_{gf}(1)=Z_{gf}(0)$ is indeed independent of $U$.
The bad news is that $Z_{gf}(0)=0$ \cite{nnogo} because, for $t=0$,
the Boltzmann weight is independent of the ghosts.
The deeper reason for this zero is likely due to Gribov copies arising
from $U(1)$ circles in the (compact) gauge group \cite{testa}.
Obviously, if we cannot non-perturbatively gauge fix pure Yang--Mills,
there is little hope of extending this to ChGTs.

Building on ref.~\cite{schaden}, we proposed
in ref.~\cite{gsnonab} to first fix only the coset $SU(N)/U(1)^{N-1}$
and then to fix the remaining $U(1)$'s without ghosts.

This procedure leads to a modified, ``equivariant"
BRST (eBRST) symmetry.  Taking $SU(2)$ for example, we first split
$V=\half A\tau_3+W^+\tau_++W^-\tau_-$, and choose the
$U(1)$-covariant gauge-fixing
$F(V)=\partial\cdot W+i[A,W]$.
We only introduce (anti-)ghosts $C^\pm$
($\Cbar^\pm$) for the generators $\tau_\pm$, which transform as
\begin{equation}
sC=(-iC^2)_{coset}\equiv -iC^2+X\,,\ \
s\Cbar=-ib\,,
\end{equation}
where also $b$ lives only in the coset, and $X=iC^2_{U(1)}
\equiv (i/2)\tau_3\tr(\tau_3C^2)$.
We thus have $s^2C=-i[X,C]=\delta_X C$, a $U(1)$ transformation
of $C$ with parameter $X$.
(For $\Psi$ in the fundamental {\it irrep},
$s\Psi = -iC\Psi$ and $s^2\Psi = -iX\Psi$.)
The equivariantly nil-potent algebra
$s^2=\delta_X$ requires $s^2\Cbar=-i[X,\Cbar]$,
and therefore the new rule
\begin{equation}
sb=is^2\Cbar=[X,\Cbar]\ne 0\,.
\label{b}
\end{equation}
eBRST is still nil-potent on any $U(1)$-invariant operator.
But, because of eq.~(\ref{b}), $\xi g^2\tr(b^2)$
is no longer eBRST invariant, and is replaced by
\begin{equation}
s\left(i\xi g^2\tr(\Cbar b)\right)=\xi g^2[\tr(b^2)+
4\Cbar^+C^+\Cbar^-C^-]\,.
\end{equation}
The new 4-ghost term changes the bad into good news;
we still have $dZ_{gf}/dt=0$, but now
\begin{eqnarray}
Z_{gf}(1)&=&Z_{gf}(0)=\\
&&\hspace{-1.2cm}\int d\phi dC d\Cbar db\,e^{-\xi g^2[\tr(b^2)+
4\Cbar^+C^+\Cbar^-C^-]}\ne 0 \,,\nonumber
\end{eqnarray}
is not equal to zero and independent of the gauge field, just as
required.  This is a rigorous result.

In order to construct a ChGT, complete gauge-fixing is needed.
The remaining step is to fix the eBRST-invariant theory, which
is now an abelian gauge theory, using a lattice version of
$(\partial\cdot A)^2$ and no ghosts, in order to avoid the
no-go theorem \cite{nnogo}.  The resulting modified
Slavnov--Taylor identities are still sufficient to prove
unitarity (to all orders) \cite{gsnonab}.  The intuitive reason is that since
the theories defined by $Z_{unfixed}$ and $Z_{eBRST}$ have
identical physical sectors, gauge fixing either one completely
should not change this.

The lattice gauge-fixing action also contains an irrelevant term,
which is needed for the uniqueness of the classical vacuum
and, hence, for the existence of a critical point near which PT is valid
\cite{ysgs,gsnonab}.  As a result, the fully gauge-fixed lattice Yang--Mills
action is not eBRST invariant.  However, since gauge invariance
is already broken by the chiral fermions and counter terms
are required anyway, this is not a new price to pay.

All these elements together lead to a complete construction
of lattice $SU(N)$ ChGTs.
Summarizing the main
features, what we have is
\begin{itemize}
\item Locality,
\item Perturbation theory near the critical point
defining the continuum target theory,
\item No fermion doublers, because of our control
over the longitudinal modes,
\item The theory accounts for fermion-number violation \cite{fnv},
\item Universality: our construction works for
any lattice fermion method without doublers in the classical
continuum limit \cite{bgsprl,bd}.
\end{itemize}
What we do not have is
\begin{itemize}
\item Exact gauge invariance (we need a finite number of
counter terms),
\item Manifest unitarity; but it is recovered at least
to all orders in perturbation theory.
\end{itemize}
In summary, our claim can be stated as follows: {\it If a certain
chiral gauge theory exists, our construction provides a valid,
non-perturbative lattice regularization.}  For details,
we refer to ref.~\cite{gsnonab}.

In the remaining space we compare our results with those
of refs.~\cite{mlab,mlnonab}.  These references
take the other route mentioned in the beginning: the
modification of chiral symmetry on the lattice, with the aim
of maintaining exact chiral gauge invariance.
To do this, one employs Ginsparg--Wilson fermions,
and left-handed fermions are defined through a
modified chiral projection in which $\gamma_5$ is replaced by
$ {\hat\gamma}_5(U)=\gamma_5(1-aD_{GW}(U))$.
Because of the $U$ dependence of ${\hat\gamma}_5$, the fermion
measure is also $U$ dependent. This leads to
an integrability condition on the space of lattice gauge fields
which has to be solved to define the theory
\cite{mlnonab}.  The current state of affairs is that an exact solution
was found for the abelian case (using admissible
gauge fields) \cite{mlab}; for the non-abelian case,
a solution was found only in perturbation theory \cite{mlpt}
(requiring an infinite number of irrelevant
counter terms), and no non-perturbative solution is known.
(The Witten anomaly can be recovered \cite{bc}.)

In the absence of a non-perturbative solution
this approach is incomplete (see also ref.~\cite{mgreview}).
As it does not involve gauge fixing, there is no control over the dangerous
``back reaction'' of the longitudinal modes, unless gauge invariance is exact!

Our approach does not rely on exact gauge invariance and,
as such, it does not ``immediately"
answer whether a certain unitary ChGT exists. But, whether it
exists or not is a dynamical issue which {\it can} be
investigated in our approach.


\begin{thebibliography}{9}
\bibitem{nn}
H.~B.~Nielsen and M.~Ninomiya,
Nucl.\ Phys.\ B {\bf 193}, 173 (1981);
Nucl.\ Phys.\ B {\bf 185}, 20 (1981)
[Erratum-ibid.\ B {\bf 195}, 541 (1982)].
\bibitem{ks}
L.~H.~Karsten and J.~Smit,
Nucl.\ Phys.\ B {\bf 183}, 103 (1981).
\bibitem{reviews}
D.~N.~Petcher,
arXiv:hep-lat/9301015;
Y.~Shamir,
arXiv:hep-lat/9509023.
\bibitem{mgreview}
M.~Golterman,
arXiv:hep-lat/0011027.
\bibitem{gsnonab}
M.~Golterman and Y.~Shamir,
arXiv:hep-lat/0404011.
\bibitem{ysgs}
Y.~Shamir,
Phys.\ Rev.\ D {\bf 57}, 132 (1998);
M.~Golterman and Y.~Shamir,
Phys.\ Lett.\ B {\bf 399}, 148 (1997).
\bibitem{rome}
A.~Borrelli {\it et al.},
Nucl.\ Phys.\ B {\bf 333}, 335 (1990).
\bibitem{bgsprl}
W.~Bock, M.~Golterman and Y.~Shamir,
Phys.\ Rev.\ Lett.\  {\bf 80}, 3444 (1998).
\bibitem{bd}
S.~Basak and A.~K.~De,
Phys.\ Rev.\ D {\bf 64}, 014504 (2001).
\bibitem{nnogo}
H.~Neuberger,
Phys.\ Lett.\ B {\bf 183}, 337 (1987).
\bibitem{testa}
M.~Testa,
Phys.\ Lett.\ B {\bf 429}, 349 (1998).
\bibitem{schaden}
M.~Schaden,
Phys.\ Rev.\ D {\bf 59}, 014508 (1999).
\bibitem{fnv}
M.~Golterman and Y.~Shamir,
Phys.\ Rev.\ D {\bf 67}, 014501 (2003).
\bibitem{mlab}
M.~L\"uscher,
Nucl.\ Phys.\ B {\bf 549}, 295 (1999).
\bibitem{mlnonab}
M.~L\"uscher,
Nucl.\ Phys.\ B {\bf 568}, 162 (2000).
\bibitem{mlpt}
M.~L\"uscher,
JHEP {\bf 0006}, 028 (2000).
\bibitem{bc}
O.~B\"ar and I.~Campos,
Nucl.\ Phys.\ B {\bf 581}, 499 (2000).
\end{thebibliography}
\end{document}